\documentclass[aps,prd,reprint,superscriptaddress,floatfix,nofootinbib,showkeys,showpacs,preprintnumbers,noeprint,nolongbibliography]{revtex4-1}

\pdfoutput=1

\usepackage{tabularx}
\usepackage{xspace}
\usepackage{listings}
\usepackage{lineno}

\usepackage[dvipsnames,svgnames]{xcolor} 
\usepackage{graphicx}
\usepackage{comment}

\usepackage{natbib}
\usepackage{hyperref}
\usepackage{amsmath}
\usepackage{enumitem}
\setlist{wide, labelwidth=!, labelindent=0pt}
\usepackage{multirow}
\hypersetup{    
  colorlinks      = true,
  linkcolor       = {blue},
  linkbordercolor = {white},
  citecolor       = {blue},
  citebordercolor = {white},
  urlcolor        = {blue},
  urlbordercolor  = {white},
}

\usepackage{lipsum}
\usepackage{aas_macros}
\usepackage{orcidlink}

\begin{document}

\title{Semianalytic model for decaying dark matter halos}

\author{Ethan~O.~Nadler\orcidlink{0000-0002-1182-3825}}
\email[]{enadler@ucsd.edu}
\affiliation{Department of Astronomy \& Astrophysics, University of California, San Diego, La Jolla, California 92093, USA}

\author{Andrew~J.~Benson\orcidlink{0000-0001-5501-6008}}
\email[]{abenson@carnegiescience.edu}
\affiliation{Carnegie Observatories, 813 Santa Barbara Street, Pasadena, California 91101, USA}

\begin{abstract}
Decaying dark matter (DDM) affects the evolution of cosmic structure relative to standard cold, collisionless, stable dark matter (CDM). We introduce a new semianalytic model for the effects of two-body DDM on halo structure and subhalo populations. In this scenario, cold parent dark matter particles decay into less massive daughter particles plus dark radiation with a lifetime comparable to the age of the Universe. Our DDM model is implemented in the open-source software \textsc{Galacticus} and accounts for heating (due to the velocity kicks imparted on daughter particles) and mass loss (due to the parent-daughter mass splitting). We show that decays flatten and reduce the amplitude of halos’ inner density profiles. These effects make DDM subhalos susceptible to tidal disruption, which we show yields a mass-dependent suppression of the subhalo mass function relative to CDM. Our predictions for DDM density profiles, velocity dispersion profiles, and subhalo populations are consistent with results from isolated and cosmological DDM N-body simulations. Thus, our model enables efficient and accurate exploration of DDM parameter space and will be useful for deriving constraints from upcoming small-scale structure observations.
\keywords{Cosmology, Dark matter, Galactic halos, Particle astrophysics}
\end{abstract}

\maketitle

\section{Introduction}
\label{sec:intro}
In the standard cold, collisionless dark matter (CDM) paradigm, the dark matter (DM) particle is assumed to be stable~\cite{Hambye:2010zb}. However, in a wide variety of particle models, the dominant component of the DM can decay into less massive particles. In the two-body decaying dark matter (DDM) scenario, a cold parent particle decays to a slightly lighter daughter particle and dark radiation \cite{Sanchez-Salcedo:2003pym,Strigari:2006jf,Peter:2010au}. Various models of supersymmetric weakly interacting massive particle (WIMP) DM feature this behavior~\cite{Pagels:1981ke,Weinberg:1982zq,Berezinsky:1991sp,Covi:1999ty,Feng:2003xh,Kaplinghat:2005sy,Allahverdi:2014bva}. DM models with extra dimensions can also feature such decays (e.g., within towers of neutrino or graviton DM states), in addition to other decay channels \cite{Abazajian:2000hw,Gonzalo:2022jac,Law-Smith:2023czn,Obied:2023clp,McKeen:2024fdl}. Two-body DDM is a minimal extension of CDM since decays to dark radiation alone and/or to Standard Model particles are generally more tightly constrained~\cite{Chen:2003gz,Gaskins:2016cha,Poulin:2016nat,Abazajian:2017tcc,DES:2020mpv,Nygaard:2020sow,Nygaard:2023gel,Bucko:2022kss,Simon:2022ftd,Lovell:2024qwb}.

Two-body DDM models that are compatible with observations of linear cosmic structure have decay lifetimes comparable to the age of the Universe~\cite{Clark:2020miy,FrancoAbellan:2021sxk,Bucko:2023eix}. As a result, DDM does not significantly alleviate the Hubble tension (as proposed by, e.g., Ref.~\cite{Vattis:2019efj}), although it can reduce the $S_8$ tension~\cite{Abdalla:2022yfr,Tanimura:2023bkh,Fuss:2024dam}. DDM impacts structure more significantly on smaller scales because, to conserve momentum, daughter particles receive velocity kicks from the decays. This effect heats DM halos with velocity dispersions near and below the kick velocity and can even unbind halos with sufficiently small escape velocities~\cite{Peter:2010jy,Wang:2014ina}. Thus, small-scale structure probes including the Lyman-$\alpha$ forest~\cite{Wang:2013rha,Fuss:2022zyt}, weak gravitational lensing~\cite{Wang:2010ma,Wang:2012eka}, the cluster mass function and mass-concentration relation~\cite{Peter:2010au,Peter:2010xe}, and Milky Way~(MW) satellite galaxies~\cite{Peter:2010sz,DES:2022doi} incisively test DDM physics.

DDM halo and subhalo modeling often relies on cosmological simulations. For example, Ref.~\cite{Peter:2010sz} fit a density profile model to DDM simulations of $25$ isolated MW-mass halos and applied it to transform the density profiles of (sub)halos from CDM merger trees into corresponding DDM predictions. Based on the enclosed masses of Sloan Digital Sky Survey MW satellites, Ref.~\cite{Peter:2010sz} used this model to place lower limits of $\sim 30~\mathrm{Gyr}$ on the decay lifetime for kick velocities spanning $20$ to $200~\mathrm{km\ s}^{-1}$. Meanwhile, Ref.~\cite{DES:2022doi} used six cosmological DDM zoom-in simulations of MW-mass halos (including several from Ref.~\cite{Wang:2014ina}) to model the suppression of DDM subhalo abundances relative to CDM. By including this suppression in a galaxy-halo connection framework based on CDM simulations, Ref.~\cite{DES:2022doi} thus placed lower limits on the DDM lifetime of $18$ and $29~\mathrm{Gyr}$ at kick velocities of $20$ and $40~\mathrm{km\ s}^{-1}$, respectively. These limits were based on the Dark Energy Survey Year 3 and Pan-STARRS1 MW satellite luminosity function~\cite{DES:2019vzn} and complemented the previous constraints based on stellar dynamics.

Because they are based on DDM simulations, such modeling techniques are limited by the parameter space that these simulations cover. For example, Ref.~\cite{Peter:2010sz} assumed that DDM density profiles are self-similar when scaled by virial parameters to extrapolate beyond the halo mass range covered by their simulations. However, as we will show, DDM halo evolution differs for halos with velocity dispersions above and below the kick velocity. Meanwhile, the subhalo mass function models from Ref.~\cite{DES:2022doi} were fit to simulations run at two kick velocity values. While the resulting constraints were conservatively extrapolated to larger kick velocities, explicitly modeling DDM subhalo populations at other points in parameter space will yield stronger limits. The ability to efficiently model DDM (sub)halos, including systems that are difficult to resolve in simulations, is particularly important as observations push to lower halo masses~\cite{Bechtol:2022koa}.

This situation strongly motivates a semianalytic model for DDM (sub)halo populations. Here, we extend the open-source \textsc{Galacticus} framework for structure and galaxy formation~\cite{Benson:2010kx} to include physical models for the effects of DDM on halos and subhalos.\footnote{\url{http://github.com/galacticusorg/galacticus}} We model both mass loss and velocity kicks; the latter can lead to the escape of sufficiently high-velocity daughter particles from a halo. We will show that velocity kick heating and escape is non-negligible for halos with velocity dispersions near the kick velocity and can dominate for even lower-mass systems.

We implement DDM physics in an adiabatic heating scheme that was previously used to model tidal heating~\cite{Pullen:2014gna,Yang:2020aqk,Benson:2022tzm,Du:2024sbt}. Unlike N-body simulations, our approach is modular and allows us to test the relative importance of velocity kicks and mass loss. In addition to two physical DDM parameters---the decay lifetime and kick velocity---our model only introduces one new parameter related to the efficiency of heating due to DDM mass loss.\footnote{In principle, the parent particle mass is also a free parameter. As described in Sec.~\ref{sec:model}, we assume that the parent is sufficiently heavy so as to remove this degree of freedom.} Without explicitly calibrating this parameter, our predictions agree with density and circular velocity profiles measured from DDM N-body simulations.

For subhalos, the effects of DDM on halo structure lead to the possibility of tidal disruption~\cite{Errani:2019sey,Errani:2022aru,Benson:2022tzm}, which remains difficult to disentangle from artificial disruption and subhalo finder incompleteness in cosmological simulations~\cite{Han:2017lpe,vandenBosch:2017ynq,vandenBosch:2018tyt,Green:2021vtc,Mansfield:2023prs,He:2024hsx}. By including our DDM heating and mass loss models in semianalytic merger trees, we efficiently generate high-resolution DDM subhalo populations that do not suffer from these effects. We will show that our subhalo population predictions agree with results from cosmological DDM simulations, where overlap exists, despite running at a small fraction of the computational cost. Thus, our model will be useful for generating subhalo population predictions and deriving constraints over a much wider range of DDM parameter space than achieved by previous studies.

This paper is organized as follows: in Sec.~\ref{sec:model}, we describe the two-body DDM model, our semianalytic treatment of its effects on halos, and our implementation of this physics in (sub)halo merger trees. In Sec.~\ref{sec:isolated}, we present our predictions for isolated DDM halo evolution, focusing on density and circular velocity profiles; in Sec.~\ref{sec:subhalos}, we study DDM subhalo populations, focusing on mass functions and radial distributions. In Sec.~\ref{sec:discussion}, we summarize and discuss our results.

Throughout, we adopt cosmological parameters that match the DDM N-body zoom-in simulations to which we compare, from Refs.~\cite{Wang:2014ina,DES:2022doi}: $h = 0.71$, $\Omega_{\rm m} = 0.266$, $\Omega_{\Lambda} = 0.734$, $\sigma_8 = 0.801$, and $n_s=0.963$. Halo masses are calculated using the virial overdensity~\cite{Bryan:1997dn}.

\section{Decaying Dark Matter Model}
\label{sec:model}

\subsection{Particle model}

We consider a two-body DDM model in which a parent particle $\chi$, of mass $m_\chi$, decays into a daughter particle $\chi'$, of mass $m_{\chi'}<m_{\chi}$, and massless dark radiation $\gamma'$:
\begin{equation}
    \chi \rightarrow \chi' + \gamma'.
\end{equation}
We assume that the parent particle is sufficiently massive such that it can be modeled as CDM; for example, supersymmetric WIMPs that behave like DDM often have $m_{\chi}\sim 1~\mathrm{GeV}$. Thus, for our purposes, the free parameters are the decay lifetime, $\tau \sim \mathcal{O}(\mathrm{Gyr})$, and the mass splitting of the decay,
\begin{equation}
    \epsilon \equiv \frac{m_\chi - m_{\chi'}}{m_\chi}.
\end{equation}

When a decay occurs, the daughter particle gains a velocity kick, $v_\mathrm{kick}$, that is proportional to the mass splitting and follows from conservation of momentum:
\begin{equation}
    v_{\mathrm{k}} \equiv \epsilon c.
\end{equation}
We consider models with $v_{\mathrm{kick}}~\sim 10~\mathrm{km\ s}^{-1}$ ($\epsilon \sim 10^{-5}$). These velocity kicks act as a heat source within DM halos. Decays also result in mass loss, both because the daughter particles are slightly less massive than the parent and because some daughter particles escape the halo due to the velocity kicks.

\subsection{Adiabatic heating framework}

To model the heat introduced by decays, we generalize the adiabatic heating model from Ref.~\cite{Pullen:2014gna}. Specifically, we decompose a halo into spherical mass shells and use conservation of energy to write
\begin{equation}
    -\frac{GM'(<r_\mathrm{f})}{2r_\mathrm{f}} = -\frac{GM(<r_\mathrm{i})}{2r_\mathrm{i}} + \Delta E_{\mathrm{tot}}(t,r_\mathrm{i}),\label{eq:heating}
\end{equation}
where $G$ is Newton's gravitational constant, $M'(<r_\mathrm{f})$ ($M(<r_\mathrm{i})$) is the final (initial) enclosed halo mass at a radius $r_\mathrm{f}$ ($r_\mathrm{i}$), and $\Delta E_{\mathrm{tot}}(t,r_\mathrm{i})$ is the specific energy injected into the shell at $r_\mathrm{i}$ up to time $t$. Reference~\cite{Pullen:2014gna} calculated the specific energy injected by tidal heating, $\Delta E_{\mathrm{tidal}}$; below, we derive the specific energy injected due to decays, $\Delta E_{\mathrm{DDM}}$. We combine these effects to calculate the total specific energy injected, as follows:
\begin{equation}
    \Delta E_{\mathrm{tot}}(t,r_\mathrm{i}) = \Delta E_{\mathrm{tidal}}(t,r_\mathrm{i}) + \Delta E_{\mathrm{DDM}}(t,r_\mathrm{i}).
\end{equation}
Note that our isolated halo models assume $\Delta E_{\mathrm{tidal}}=0$. 

At radii where shell crossing does not occur, mass conservation implies $M'(<r_\mathrm{f})=M(<r_\mathrm{i})$, and we can solve Eq.~(\ref{eq:heating}) for $r_\mathrm{f}$. Performing this procedure over a grid of $r_\mathrm{i}$ yields the mass profile after heat injection. However, as noted by Ref.~\cite{2024PhRvD.110b3019D}, shell crossing \emph{can} result from tidal heating for density profiles with shallow logarithmic slopes, i.e., $\rho(r) \propto r^{-\alpha}$ and $\alpha < 1$, which is particularly relevant for our DDM models. To approximate the effects of shell crossing, which occurs when $\mathrm{d}r_{\mathrm{f}}/\mathrm{d}r_{\mathrm{i}}<0$, Ref.~\cite{2024PhRvD.110b3019D} solves for the shell-crossing radius, $r_{\mathrm{c}}$, using
\begin{equation}
    \frac{\mathrm{d}}{\mathrm{d}r}\left(\frac{\Delta E_{\mathrm{tot}}(t,r)}{GM(<r)/2r}\right)\Bigg\lvert_{r=r_{\mathrm{c}}} = 0,\label{eq:monotonic_condition}
\end{equation}
and fixes the heating energy ratio to $\Delta E_{\mathrm{tot}}(t,r_{\mathrm{c}})/(GM(<r_{\mathrm{c}})/2r_{\mathrm{c}})$ for $r<r_{\mathrm{c}}$. In our DDM model, shell crossing always occurs at large radii according to this criterion. Instead, we determine $r_{\mathrm{c}}$ using the weaker condition
\begin{equation}
    \left(\frac{\mathrm{d}r_{\mathrm{f}}}{\mathrm{d}r_{\mathrm{i}}}\right)\Bigg\lvert_{r=r_{\mathrm{c}}} = 0,\label{eq:monotonic_weak}
\end{equation}
and we use the fixed heating energy ratio within $r_{\mathrm{c}}$ described above. This procedure ensures that shell crossing is captured, but introduces a feature in our density profile predictions at $r_{\mathrm{c}}$, as discussed below.

\subsection{DDM retained fraction}

Before calculating $\Delta E_{\mathrm{DDM}}$, it is useful to calculate the mass lost due to decays, and to find the kick energy deposited in the halo by daughter particles which \emph{do not} escape the profile. We begin by assuming that the initial distribution of particle velocities is a Maxwell-Boltzmann distribution, truncated at the escape velocity. We then compute the mean specific energy of retained particles minus their mean specific energy before decays.

Specifically, the truncated Maxwell-Boltzmann distribution is
\begin{equation}
  p(v,\theta|r,s) = \left\{ \begin{array}{ll} A^{-1} v^2 \exp\left(-\frac{1}{2}\left[\frac{v}{s}\right]^2\right) & \hbox{if } v < v_\mathrm{e}(r) \\ 0 &  \hbox{if } v \ge v_\mathrm{e}(r), \end{array} \right.
\end{equation}
where $s$ is the velocity width, $v_\mathrm{e}(r)$ is the local escape velocity, $v$ is the particle speed, $\theta$ is the direction of
the particle velocity relative to the $z$-axis, and
\begin{equation}
  A(r|s) = \sqrt{2 \pi} s^3 \hbox{erf}\left( \frac{v_\mathrm{e}(r)}{\sqrt{2}s} \right) - 2 v_\mathrm{e} s^2 \exp \left( -\frac{1}{2}\left[\frac{v_\mathrm{e}(r)}{s}\right]^2\right)
\end{equation}
is a radially-dependent normalization factor. We fix $s$ by requiring that the root-mean-square particle velocity is $\sqrt{3}$ times the one-dimensional velocity dispersion, $\sigma_v$. We estimate $\sigma_v$ by solving the Jeans equation in the original profile, assuming an isotropic velocity distribution. We assume an isotropic velocity distribution, such that $p(v,\theta|r,s)$ is independent of $\theta$.

\begin{figure}[t!]
 \hspace{-2mm}
 \includegraphics[width=\columnwidth]{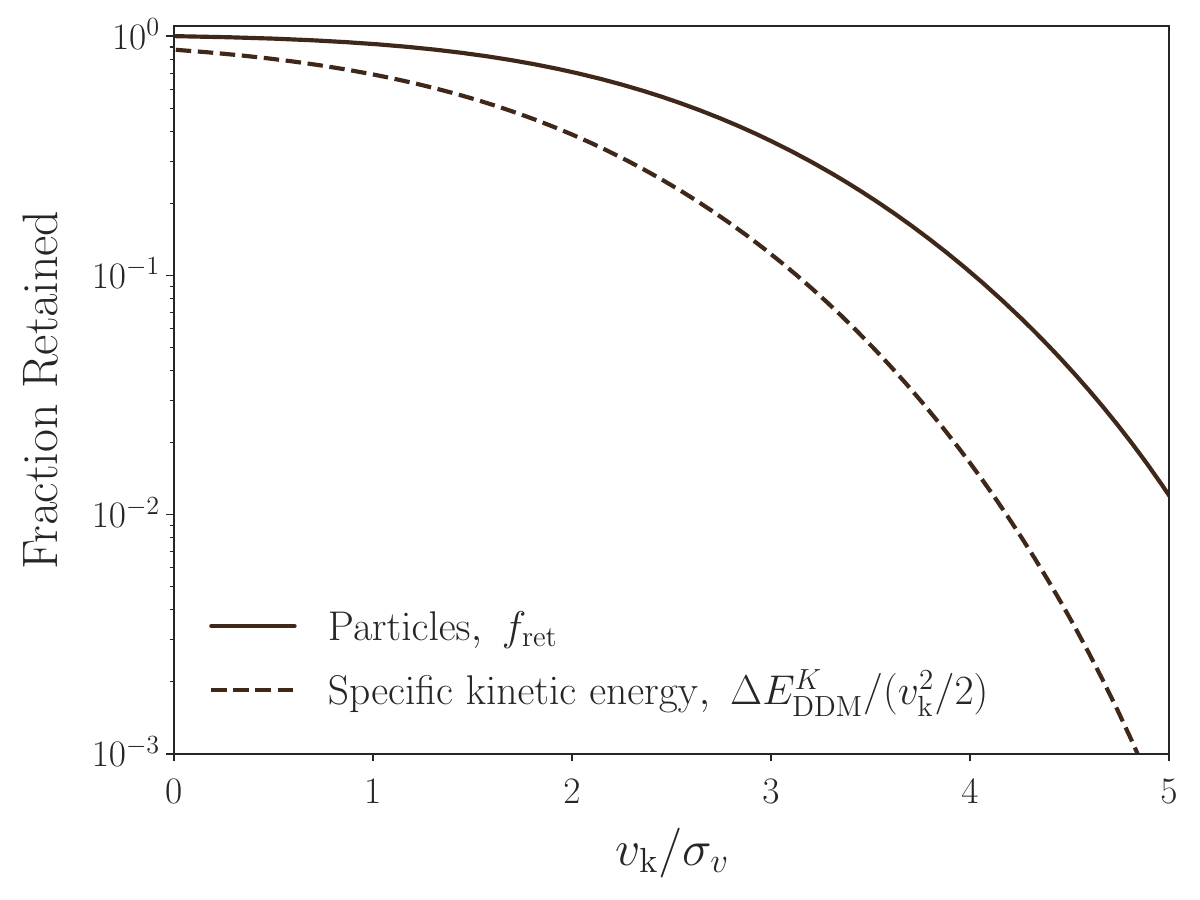}
 \vspace{-2.75mm}
 \caption{Fraction of decayed particles (solid) and mean specific kinetic energy (in units of $v_\mathrm{k}^2/2$; dashed) retained in an isolated DDM halo. Retained fractions are plotted as a function of the kick velocity, $v_{\mathrm{k}}$, normalized to the local velocity dispersion, $\sigma_v$. Results are shown for an escape velocity $v_{\mathrm{e}}=3\sigma_v$, characteristic of the inner regions of CDM halos.} \label{fig-retained}
\end{figure}

Assuming, without loss of generality, that the kick is along the positive $z$-axis, the instantaneous fraction of decayed particles that are retained is
\begin{align}
  f_\mathrm{ret}(r) =  \int_{-1}^{+1} \mathrm{d}&\cos\theta \int_0^{v_\mathrm{e}} \mathrm{d}v\Big[p(v,\theta|r,s) \nonumber\\ & \times H\left( v^2 + v_\mathrm{k}^2 + 2 v v_\mathrm{k} \cos\theta < v_\mathrm{e}^2(r) \right)\Big],
\end{align}
where $H(x) = 1$ if $x$ is true and 0 otherwise, and $\theta$ is the polar angle. Solving the inequality for the velocity that will remain bound as a function of $\theta$ gives two solutions\footnote{Note that there can be a minimum initial speed required for the particle to remain bound \emph{if} the kick has a significant component antiparallel to the initial velocity. In such cases the initial velocity partially cancels some of the kick velocity, leaving the particle below the escape velocity after the decay.}: $v_\mathrm{max|min}(r,\theta) = \hbox{max}\left(0,\pm \left[v^2_\mathrm{e}(r) - v_\mathrm{k}^2 \sin^2\theta \right]^{1/2} - v_\mathrm{k} \cos
\theta\right)$. Thus,
\begin{equation}
  f_\mathrm{ret}(r) =  \int_{-1}^{+_1} \mathrm{d}\cos\theta \int_{v_\mathrm{min}(r,\theta)}^{v_\mathrm{max}(r,\theta)} \mathrm{d}v ~p(v,\theta|r,s).
  \label{eq:decayingDMRetainedFraction}
\end{equation}
Note that the fraction of particles that escape is simply $f_{\mathrm{esc}}(r)=1-f_{\mathrm{ret}}(r)$. The solid line in Fig.~\ref{fig-retained} shows the retained fraction as a function of $v_{\mathrm{k}}$, assuming $v_\mathrm{e}=3\sigma_v$. As $v_{\mathrm{k}}/\sigma_v$ increases, more particles are able to escape from the halo, resulting in a lower retained fraction.

Given this instantaneous retained fraction, the integrated mass fraction retained in a halo at radius $r$ and time $t$ is given by
\begin{equation}
    f_{\mathrm{ret}}(t,r) = [1-f_\mathrm{d}(t)]+ [f_\mathrm{d}(t) (1-\epsilon) f_\mathrm{ret}(r)],\label{eq:fr}
\end{equation}
where $f_\mathrm{d}(t) \equiv 1-\exp(-t/\tau)$ is the fraction of particles that have undergone decays up to time $t$. In Eq.~\ref{eq:fr}, the first term represents the particles that have not yet decayed, while the second represents the particles that have decayed but have not escaped from the halo.

\subsection{DDM energy injection}

We model $\Delta E_{\mathrm{DDM}}$ as the sum of two effects: specific energy injected due to velocity kicks, and specific energy injected due to mass loss. We account for these effects separately from overall mass reduction due to decays, which we treat in the following subsection.

We calculate the specific energy injected due to decays as follows: first, the newly-created daughter particles gain kinetic energy. Naively,
the specific kinetic energy added to the halo by decays up to time $t$ is proportional to
$f_\mathrm{d}(1-f_{\mathrm{esc}}) v_{\mathrm{kick}}^2/2$, i.e., the fraction of particles that have decayed and are retained in the halo times the specific kinetic energy imparted by each decay.

More precisely, we follow the same approach used for the retained particle fraction to compute the mean specific kinetic energy change of retained daughter particles,
\begin{align}
  &\Delta E_{\mathrm{DDM}}^{K}(r) =  \int_{-1}^{+1} \mathrm{d}\cos\theta \int_0^{v_\mathrm{e}} \mathrm{d}v \Big[p(v,\theta|r,s)&\nonumber \\ & \times \frac{1}{2} \left( v_\mathrm{k}^2 + 2 v v_\mathrm{k} \cos\theta \right) \times H\left( v^2 + v_\mathrm{k}^2 + 2 v v_\mathrm{k} \cos\theta < v_\mathrm{e}^2(r) \right)\Big],&\label{eq:delta_e_unsimplified}
\end{align}
where the specific kinetic energy term in the velocity integrand does not contain $v^2$ because this cancels when the original specific kinetic energy is subtracted. Eq.~(\ref{eq:delta_e_unsimplified}) simplifies to
\begin{align}
  \Delta E_{\mathrm{DDM}}^{K}(r) =  \int_{-1}^{+1} \mathrm{d}\cos\theta &\int_{v_\mathrm{min}(r,\theta)}^{v_\mathrm{max}(r,\theta)} \mathrm{d}v\Big[p(v,\theta|r,s)&\nonumber \\ &\times \frac{1}{2} \left( v_\mathrm{k}^2 + 2 v v_\mathrm{k} \cos\theta \right)\Big].&
  \label{eq:decayingDMRetainedEnergy}
\end{align}
The dashed line in Fig.~\ref{fig-retained} shows the mean retained specific kinetic energy as a fraction of daughter particles' specific kinetic energy, $v_\mathrm{k}^2/2$. Note that the retained energy fraction is always lower than the retained particle fraction. This is because particles that had higher energies prior to decay are more likely to escape, such that retained particles are biased toward lower prekick energies.

Next, we calculate the change in the specific gravitational binding energy due to mass loss,
\begin{align}
    \Delta E_{\mathrm{DDM}}^U(r) &= \gamma \frac{G\Delta M(<r)}{r}&\nonumber \\ &= \gamma\left[ (1-f_{\mathrm{ret}}(r)) + \epsilon f_{\mathrm{ret}}(r) \right]\frac{GM(<r)}{r},&\label{eq:delta_E_U_DDM}
\end{align}
where the first term in the square brackets represents the fraction of mass directly kicked out of the profile, and the second term represents the reduction in mass of the decay products that are retained. Here, $\gamma$ parametrizes the efficiency of energy injection due to mass loss and $M(<r)$ is the mass initially enclosed within radius $r$. We set $\gamma=0.5$ for our fiducial results; for this value, assuming virial equilibrium such that the energy of the shell is half of its gravitational binding energy, the energy of the shell reaches zero only when the fraction of particles that decay and are not retained reaches unity. We study the effects of varying $\gamma$ in Appendix~\ref{sec:gamma}. Note that we apply the mass loss itself in a subsequent step described in Sec.~\ref{sec:direct_mass_loss}.

Finally, we add the velocity kick and mass loss energy injection to calculate the specific energy injected up to time $t$,
\begin{equation}
    \Delta E_{\mathrm{DDM}}(t,r) = f_\mathrm{d}(t)\left[\Delta E_{\mathrm{DDM}}^{K}(r)+\Delta E_{\mathrm{DDM}}^{U}(r)\right].
\end{equation}

\subsection{Direct effect of mass loss}
\label{sec:direct_mass_loss}

In addition to including the energy injected by mass loss in our heating calculations, we directly include the overall reduction in mass due to decays by requiring that DDM halo density profiles satisfy
\begin{equation}
    \rho(t,r) = f_{\mathrm{ret}}(t,r)\rho(t=0,r),\label{eq:M_t}
\end{equation}
where $\rho(t=0,r)$ is the initial density profile. We implement Eq.~(\ref{eq:M_t}) as a radially dependent scaling of DDM halo density profiles, as a function of time. Note that this effect is always present in our calculations, even when mass loss--induced heating is switched off.

\begin{figure*}[t!]
\centering
\includegraphics[width=0.49\textwidth]{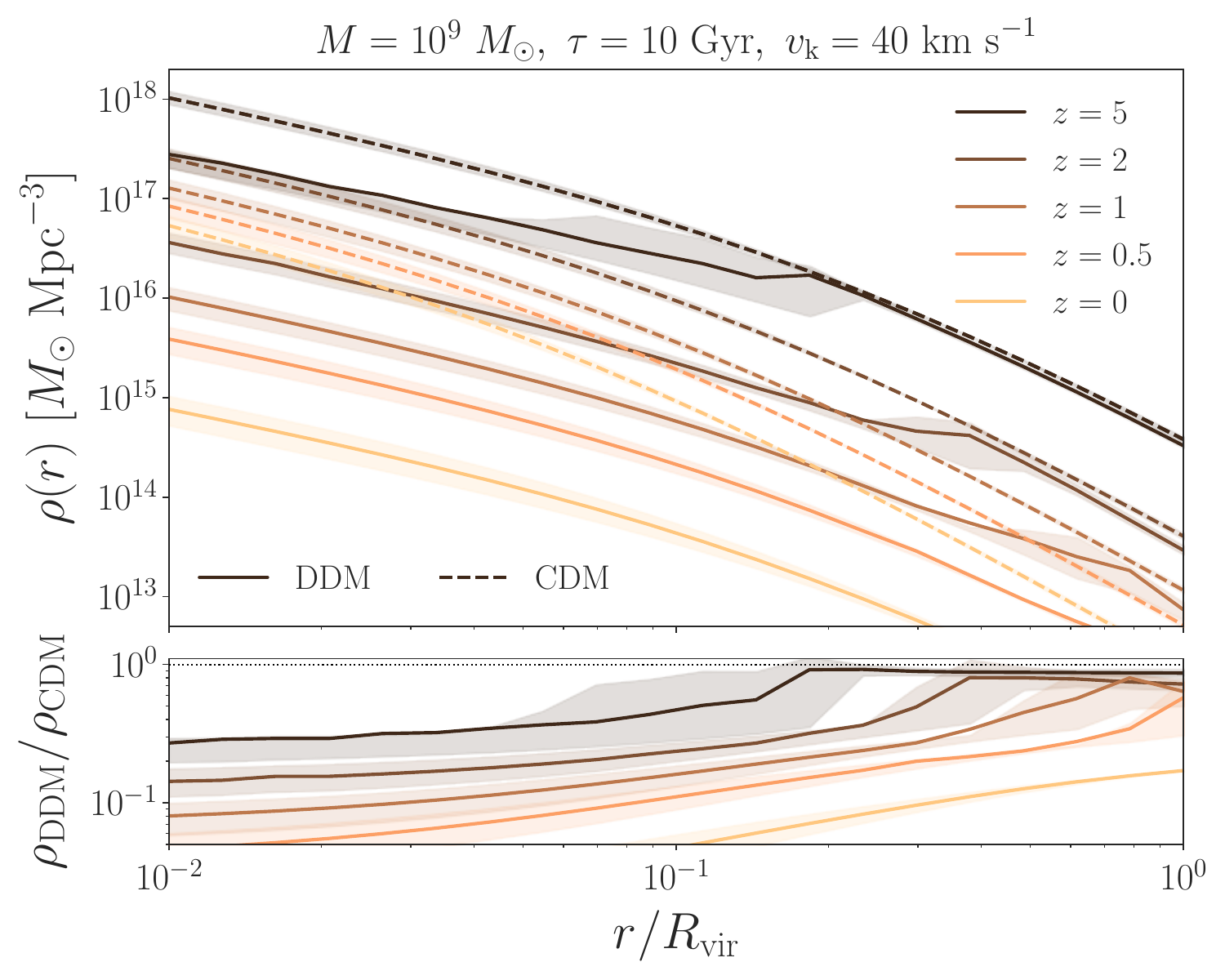}
\includegraphics[width=0.49\textwidth]{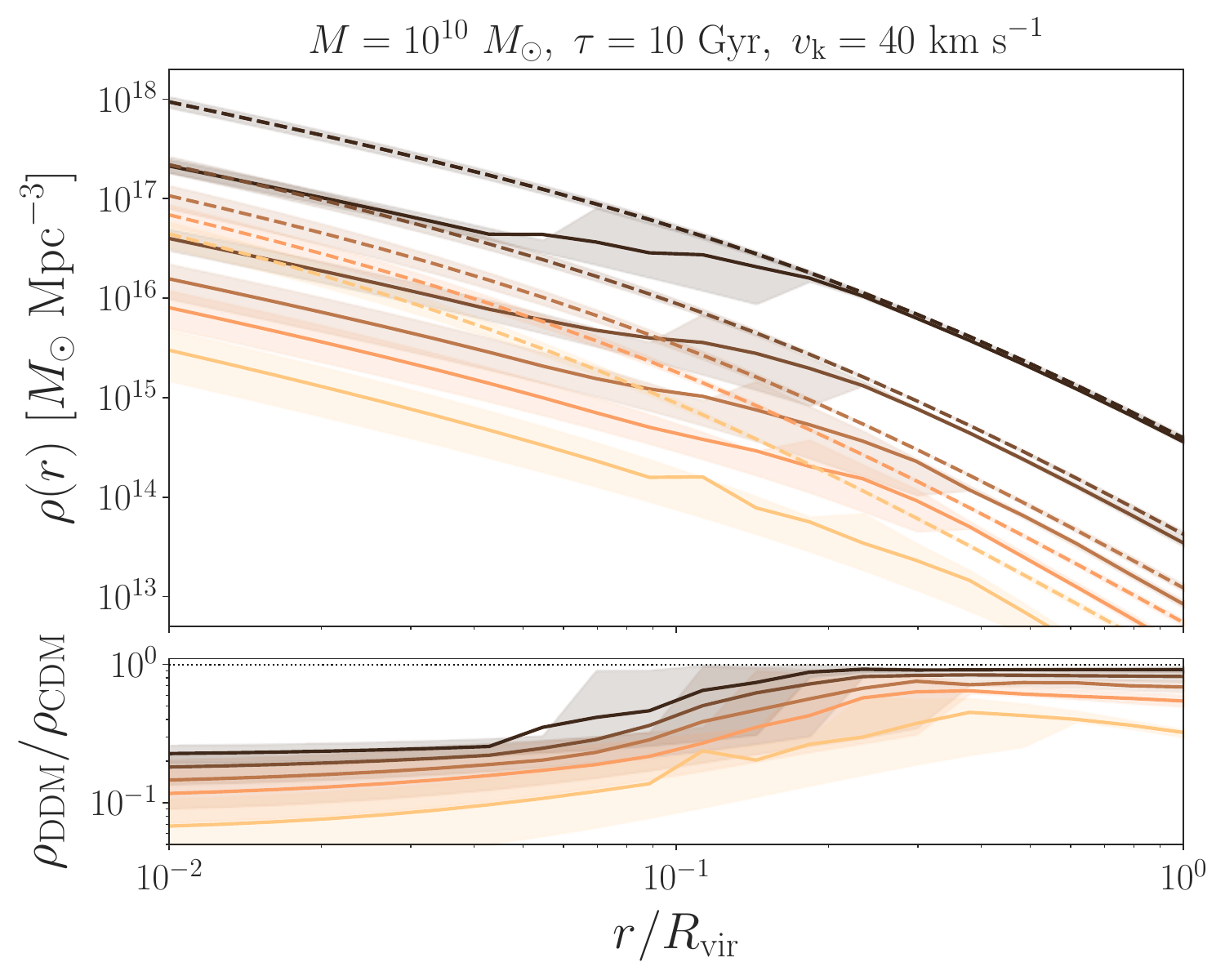}\vspace{-2mm}
\includegraphics[width=0.49\textwidth]{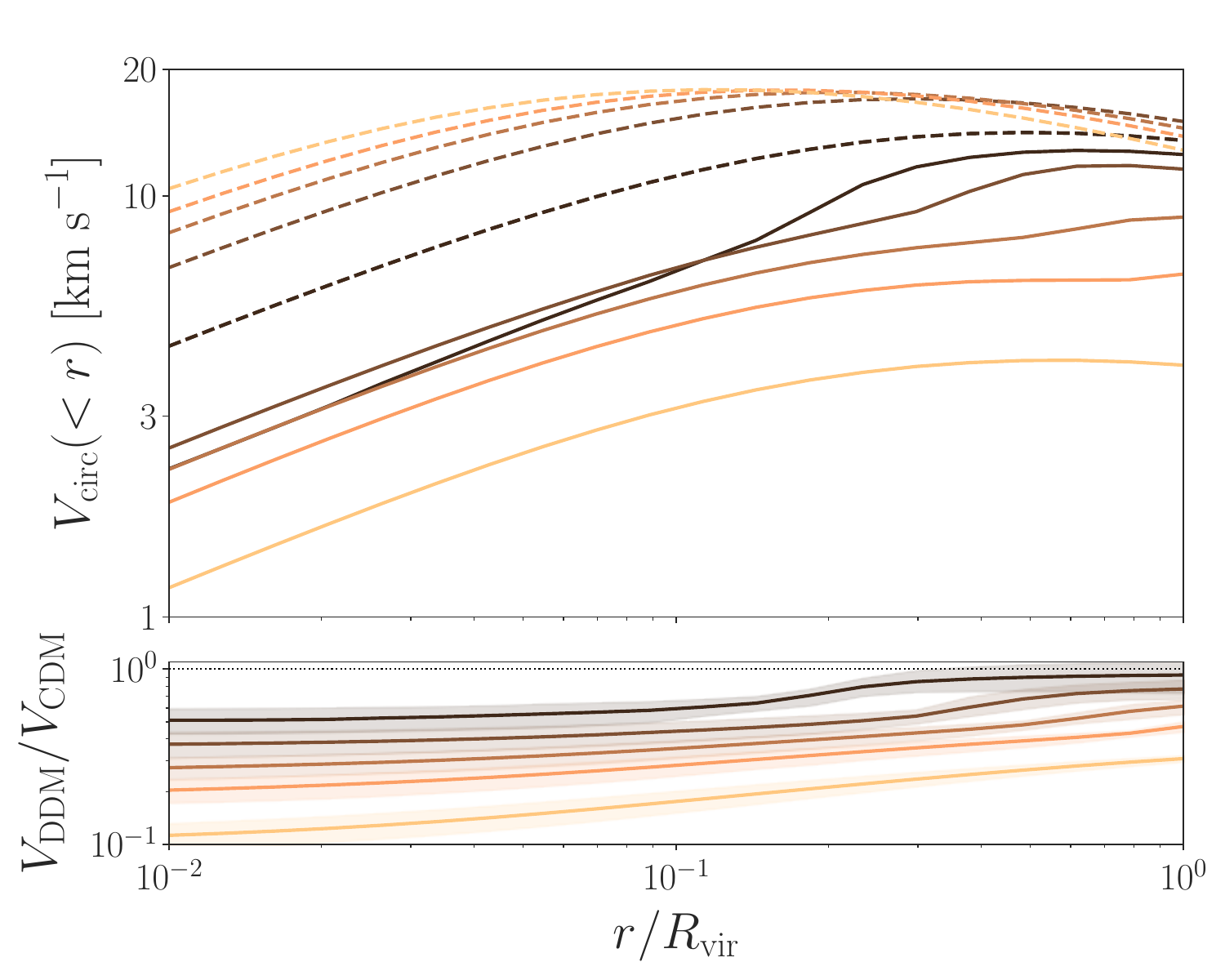}
\includegraphics[width=0.49\textwidth]{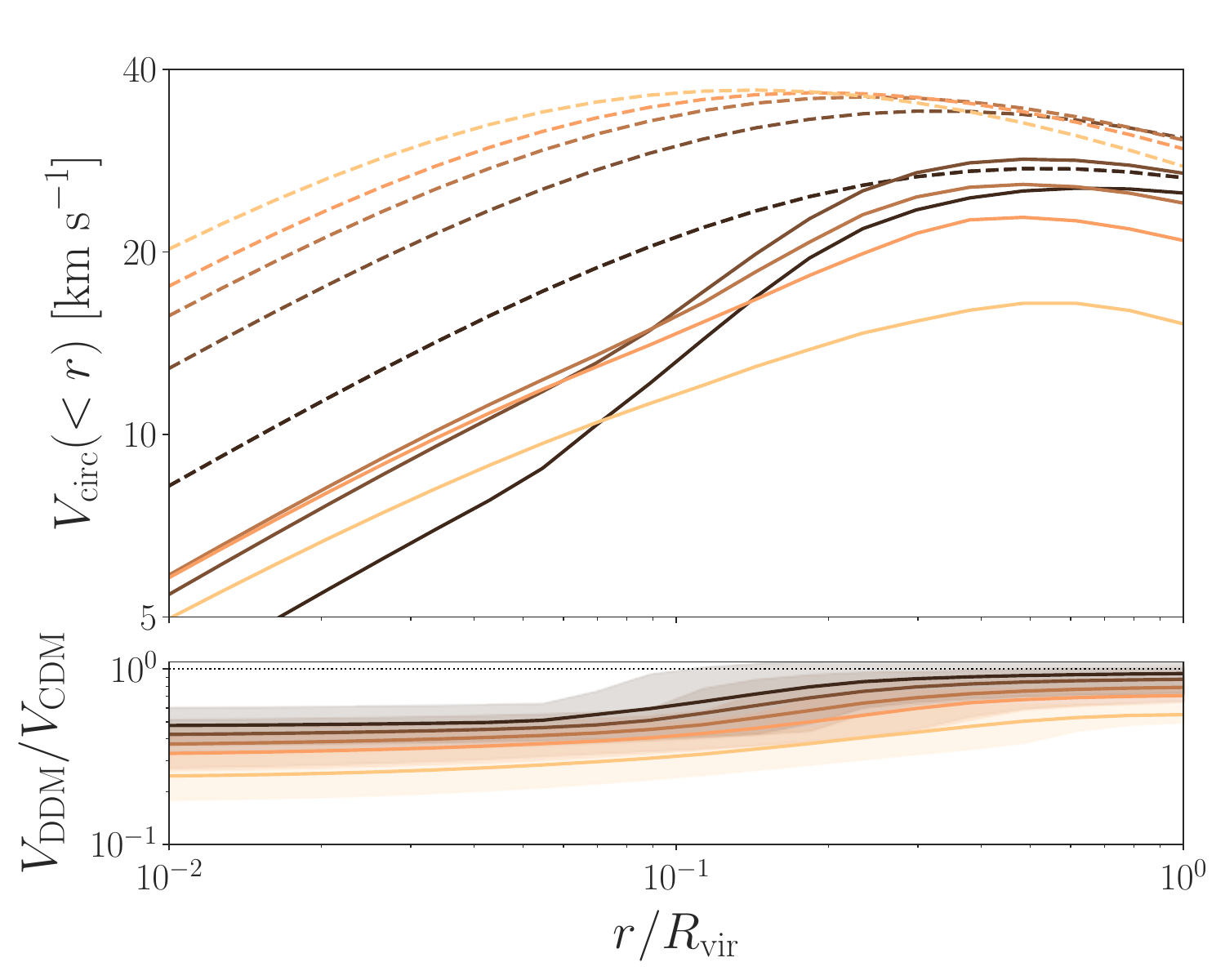}
\vspace{-2mm}
    \caption{Density (top) and circular velocity (bottom) profiles for a population of $100$ isolated halos with $z=0$ masses of $10^{9}~M_{\mathrm{\odot}}$ (left) and $10^{10}~M_{\mathrm{\odot}}$ (right). Solid lines show mean profiles in a DDM model with $\tau=10~\mathrm{Gyr}$ and $v_{\mathrm{k}}=40~\mathrm{km\ s}^{-1}$. From dark to light, profiles are evaluated at $z=5$, $2$, $1$, $0.5$, and $0$. Bands show $16\%$ to $84\%$ percentiles and are omitted from the circular velocity profiles for visual clarity. Bottom panels show ratios of DDM to CDM profiles.}
    \label{fig:density_velocity}
\end{figure*}

\section{Isolated Halos}
\label{sec:isolated}

We begin by studying the effects of decays on density and circular velocity profiles of isolated halos, where the circular velocity is defined as $V_{\mathrm{circ}}(<r)\equiv\sqrt{GM(<r)/r}$. When calculating the DDM retained fraction in these runs and for our subhalo population calculations in Sec.~\ref{sec:subhalos}, we determine each halo's escape velocity profile directly from its gravitational potential, rather than assuming $v_{e}=3\sigma_v$ as in Fig.~\ref{fig-retained}. Throughout this section, we initialize DDM halos based on the $z=0$ masses they would have in the absence of mass loss to ensure a fair comparison with CDM.

\subsection{Setup}

For illustration, we choose a DDM model with $\tau=10~\mathrm{Gyr}$ and $v_{\mathrm{k}}=40~\mathrm{km\ s}^{-1}$, which is ruled out by current MW satellite analyses~\cite{Peter:2010sz,DES:2022doi} and thus expected to affect halos with masses of $\sim 10^8$ to $10^{11}~M_{\mathrm{\odot}}$ that host dwarf galaxies~\cite{Bullock:2017xww,DES:2019ltu}. We consider isolated halos with $z=0$ masses of $M=10^9$ and $10^{10}~M_{\mathrm{\odot}}$, which have maximum circular velocities, $V_{\mathrm{max}}$, of $\approx 20$ and $40~\mathrm{km\ s}^{-1}$ at $z=0$, respectively. We expect the $10^9~M_{\mathrm{\odot}}$ halo to be more significantly affected by decays because it has $V_{\mathrm{max}}<v_{\mathrm{k}}$, which enables a significant fraction of daughter particles to escape and implies that kick velocity and mass loss--induced heating are more effective.

To evolve these halos, we construct merger trees using the algorithm from Ref.~\cite{Parkinson:2007yh}, as implemented in \textsc{Galacticus} by Ref.~\cite{Benson:2012su}. We apply the concentration model from Ref.~\cite{Ludlow:2016ifl} to predict the formation histories of
halos with resolved progenitors, with the best-fit parameters from Ref.~\cite{Benson181206026}, and we apply the concentration-mass relation from Ref.~\cite{Diemer:2018vmz} with a scatter of
$0.16~\mathrm{dex}$ for halos without progenitors. For all merger trees we present in this section, progenitors are resolved to a fractional resolution of $10^{-4}$ times the final host halo mass.

\subsection{Density and circular velocity profile evolution}

Density and circular velocity profiles for a population of $100$ isolated halos are shown in Fig.~\ref{fig:density_velocity} for the $10^9$ (left) and $10^{10}~M_{\mathrm{\odot}}$ (right) cases. At both masses, DDM clearly affects the profiles relative to CDM, with more pronounced effects at late times.\footnote{At low redshifts, the CDM profiles in Fig.~\ref{fig:density_velocity} mainly change due to pseudoevolution, since we normalize radii to $R_{\mathrm{vir}}(z)$~\cite{Diemer:2012mw}.} At small radii, DDM reduces the amplitude of and slightly flattens density profiles due to the mass loss and heating effects discussed above. Note that DDM does not produce prominent density cores, consistent with the lack of an isothermal region in the inner circular velocity profiles. These effects are all consistent with previous simulations of isolated DDM halos, and we have checked that our predictions are consistent with the density and velocity dispersion profiles from Ref.~\cite{Peter:2010jy}. Finally, we note that the slight discontinuity in our predicted DDM profiles at larger radii occurs at the shell-crossing radius determined by Eq.~\ref{eq:monotonic_weak}.

There is a striking difference between the evolution of the $10^9$ and $10^{10}~M_{\mathrm{\odot}}$ halos shown in Fig.~\ref{fig:density_velocity}. In particular, the $10^9~M_{\mathrm{\odot}}$ halo unbinds shortly after $z=1$ in all realizations. It is also significantly more affected at $z=2$ and $5$ compared to its $10^{10}~M_{\mathrm{\odot}}$ counterpart, even when normalizing to the respective CDM profiles, which is clear from the bottom residual panels. These results confirm that the evolution of halos with large characteristic internal velocities, $V_{\mathrm{max}}\gtrsim v_{\mathrm{k}}$, qualitatively differs from that of halos with $V_{\mathrm{max}}\lesssim v_{\mathrm{k}}$.

At fixed halo mass, the efficacy of DDM heating and mass loss correlates with initial concentration. This is expected since amount of heating energy injected depends on the underlying gravitational potential. In particular, low-concentration halos have shallower potential wells, allowing a larger fraction of daughter particles to escape, and vice versa for high-concentration halos. This dependence is reflected in Fig.~\ref{fig:density_velocity} by increased halo-to-halo scatter in DDM relative to CDM.

\subsection{$R_{\mathrm{max}}$--$V_{\mathrm{max}}$ relation}

To capture DDM effects on halo profiles over a range of halo masses and models, we now study the relation between $V_{\mathrm{max}}$ and the radius within a halo at which it is achieved, $R_{\mathrm{max}}$. The $R_{\mathrm{max}}$--$V_{\mathrm{max}}$ relation contains the same information as the mass-concentration relation for Navarro-Frenk-White (NFW; \cite{Navarro:1996gj}) halos, while generalizing to non-NFW profiles in nonstandard DM models, including self-interacting dark matter (SIDM; \cite{Yang:2022mxl}).

\begin{figure*}[t!]
\includegraphics[angle=270,origin=c,width=\textwidth,trim={-0.5cm 0cm 6cm 0.5cm}]{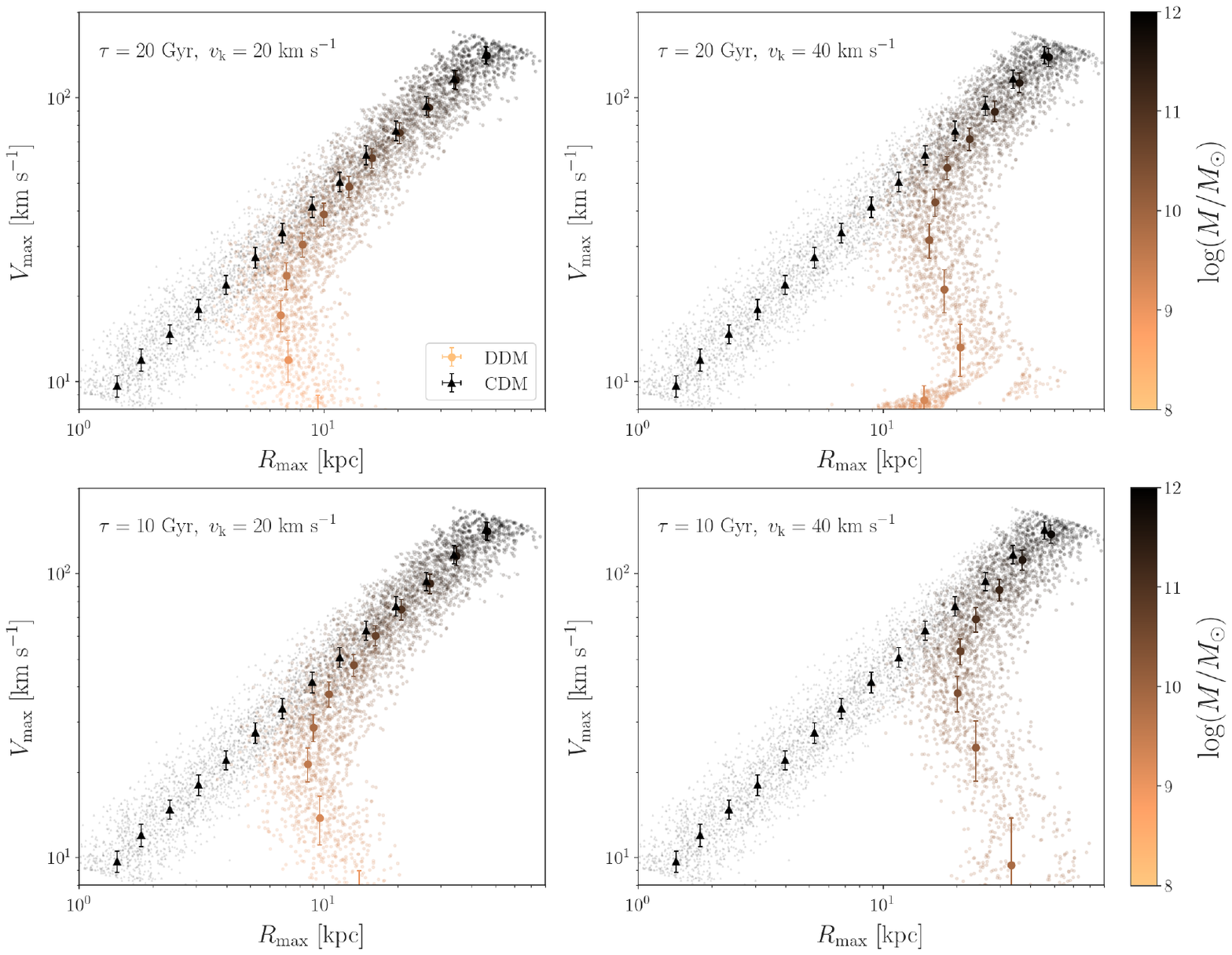}
\vspace{-2mm}
    \caption{$R_{\mathrm{max}}$--$V_{\mathrm{max}}$ relation at $z=0$ from $10^3$ isolated halo merger trees in CDM (black triangles) and four DDM models (colored circles) with $\tau=20~\mathrm{Gyr}$ and $v_{\mathrm{k}}=20~\mathrm{km\ s}^{-1}$ (top left), $\tau=20~\mathrm{Gyr}$ and $v_{\mathrm{k}}=40~\mathrm{km\ s}^{-1}$ (top right), $\tau=10~\mathrm{Gyr}$ and $v_{\mathrm{k}}=20~\mathrm{km\ s}^{-1}$ (bottom left), and $\tau=10~\mathrm{Gyr}$ and $v_{\mathrm{k}}=40~\mathrm{km\ s}^{-1}$ (bottom right). The small markers show individual halos, and the large markers and error bars show the mean and $1\sigma$ spread of $V_{\mathrm{max}}$ measurements in bins of $R_{\mathrm{max}}$. DDM points are colored by $z=0$ halo mass.}
    \label{fig:rmax_vmax}
\end{figure*}

Figure~\ref{fig:rmax_vmax} compares the $R_{\mathrm{max}}$--$V_{\mathrm{max}}$ relation at $z=0$ from $10^3$ isolated halo merger trees in CDM and four DDM models with $\tau\in [10,~ 20]~\mathrm{Gyr}$ and $v_{\mathrm{k}}\in [20~, 40]~\mathrm{km\ s}^{-1}$. We generate these merger trees using the same settings as Fig.~\ref{fig:density_velocity}. At high masses, $M\gtrsim 10^{11}~M_{\mathrm{\odot}}$ ($V_{\mathrm{max}}\gtrsim 100~\mathrm{km\ s}^{-1}$), the $R_{\mathrm{max}}$--$V_{\mathrm{max}}$ relations are similar in CDM and DDM, although the DDM relations are slightly shifted toward lower $V_{\mathrm{max}}$ and larger $R_{\mathrm{max}}$. This is mainly due to the profile heating effects discussed above. These shifts become more severe at lower masses, causing DDM halos to ``peel off'' the CDM relation. In particular, low-mass halos are pushed toward very large $R_{\mathrm{max}}$ given their $V_{\mathrm{max}}$.

The location of this feature in the DDM $R_{\mathrm{max}}$--$V_{\mathrm{max}}$ relation, and the halo masses it affects, are determined by the underlying model parameters. In particular, DDM models with smaller $\tau$ and larger $v_{\mathrm{k}}$ impact density and circular velocity profiles more significantly at a given halo mass (see Fig.~\ref{fig:density_velocity}), due to both heating and daughter particle escape. As a result, the appearance of the bend in the DDM $R_{\mathrm{max}}$--$V_{\mathrm{max}}$ relation shifts toward larger halo masses, i.e., toward the upper right in each panel of Fig.~\ref{fig:rmax_vmax}, for more extreme models. These DDM models also unbind low-mass halos more efficiently, which explains why halos in the ``tail'' of this bend in the $R_{\mathrm{max}}$--$V_{\mathrm{max}}$ relation have systematically larger masses (corresponding to darker colors) in the more extreme DDM models. For example, see the top right and particularly the bottom right panel of Fig.~\ref{fig:rmax_vmax}. Note that some halos in this tail are impacted by the discontinuity in our predicted profiles due to shell crossing, which creates an apparent bifurcation in the $R_{\mathrm{max}}$--$V_{\mathrm{max}}$ relation for some DDM models (e.g., see the $\tau=20~\mathrm{Gyr}$, $v_{\mathrm{k}}=40~\mathrm{km\ s}^{-1}$ in the top right panel of Fig.~\ref{fig:rmax_vmax}).

\begin{figure*}[t!]
\centering
\includegraphics[width=0.49\textwidth]{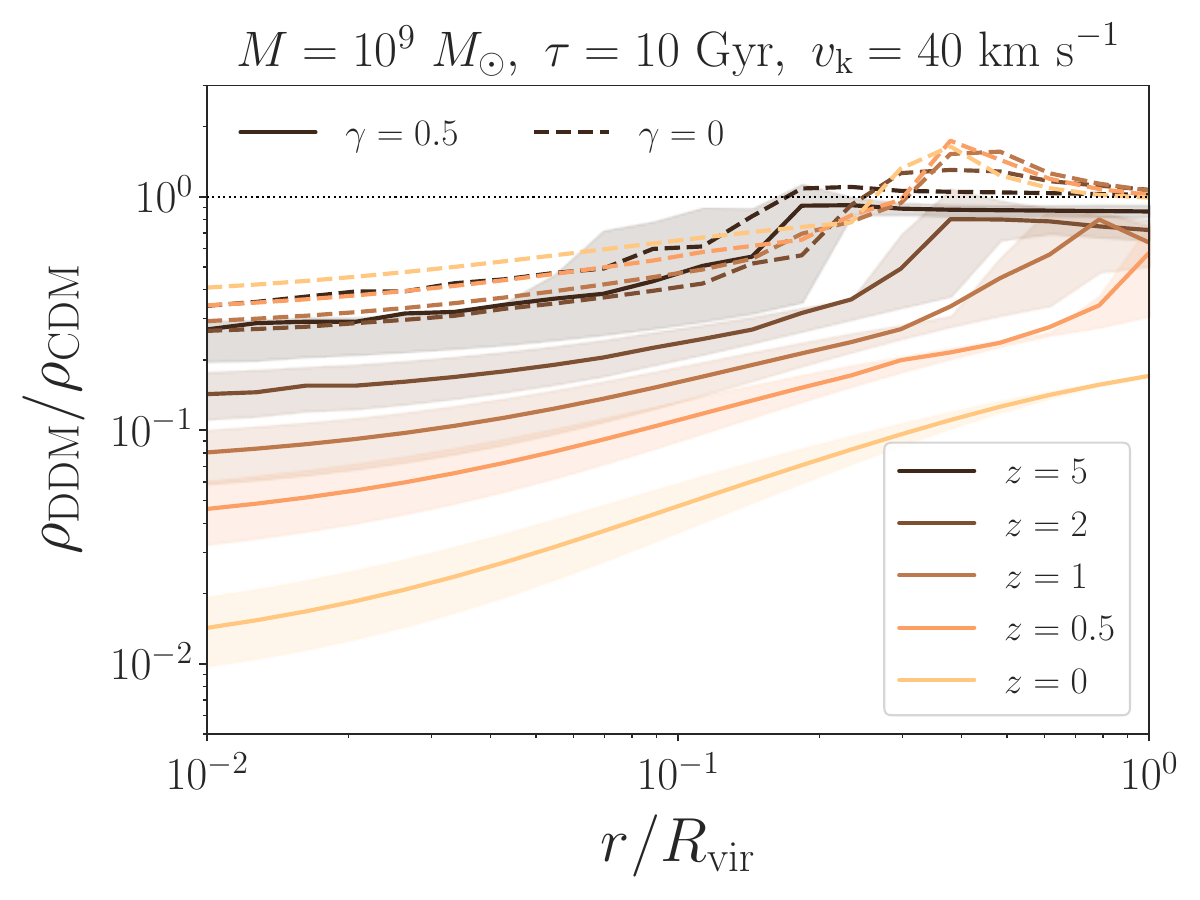}
\includegraphics[width=0.49\textwidth]{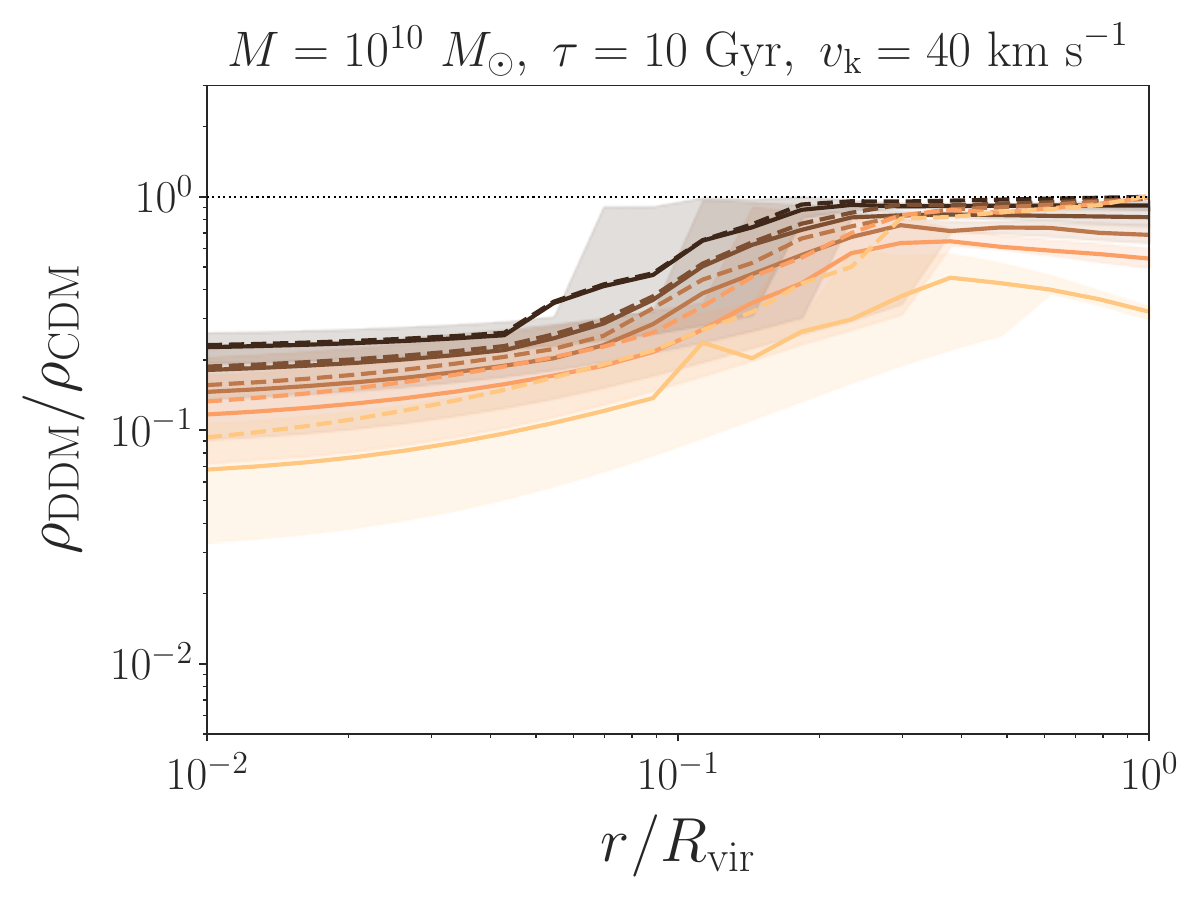}
\vspace{-2mm}
    \caption{Ratio of DDM to CDM density profiles for a DDM model with $\tau=10~\mathrm{Gyr}$ and $v_{\mathrm{k}}=40~\mathrm{km\ s}^{-1}$, for a DDM model with only velocity kick heating ($\gamma=0$; dashed) and for our fiducial DDM implementation with both velocity kick and mass loss-induced heating ($\gamma=0.5$; solid). Lines, bands, and colors otherwise follow the format of Fig.~\ref{fig:density_velocity}.}
    \label{fig:density_comparison}
\end{figure*}

We briefly contrast these results with the effects of SIDM on halo structure. Using cosmological DM-only N-body simulations of a strong, velocity-dependent SIDM model, Ref.~\cite{Yang:2022mxl} showed that, for SIDM halos in the core-forming phase, $R_{\mathrm{max}}$ increases but $V_{\mathrm{max}}$ does not significantly decrease. Meanwhile, core-collapsing SIDM halos shift toward significantly lower $R_{\mathrm{max}}$ and somewhat higher $V_{\mathrm{max}}$. The former (coring) shift is qualitatively similar to but less extreme than DDM; even for (sub)halos below the resolution limit of cosmological simulations, SIDM studies do not report bends in the $R_{\mathrm{max}}$--$V_{\mathrm{max}}$ relation like those shown for DDM in Fig.~\ref{fig:rmax_vmax} (e.g., see Ref.~\cite{Ando:2024kpk}). The latter (core-collapsing) shift is qualitatively distinct from DDM, implying that the models can be distinguished with sufficiently precise inferences of halo rotation curves.

\subsection{Effects of velocity kicks versus mass loss}

We now study the effects of DDM mass loss, velocity kicks, and mass loss heating in succession. First, we run isolated merger trees following the procedure above for a DDM model that reduces halo masses due to the parent-daughter mass splitting only, and does not include heating due to either velocity kicks or mass loss. The resulting DDM density profiles are nearly identical to CDM in the inner regions, where the escape fraction is very small (see Fig.~\ref{fig-retained}). In the outer regions, DDM density profiles are slightly suppressed. For example, $\rho_{\mathrm{DDM}}/\rho_{\mathrm{CDM}}$ approaches $\approx 0.75$ at $z=0$ for a DDM model with $\tau=10~\mathrm{Gyr}$ and $v_{\mathrm{k}}=40~\mathrm{km\ s}^{-1}$; this is consistent with the expected mass reduction of $(1-\epsilon)\times [1-\exp(-13.8~\mathrm{Gyr}/\tau)]=0.748$ in this model. Thus, DDM halos' outer density profiles are reduced by a relatively small factor due to mass loss alone, while their inner mass distributions are unaffected.

Next, Fig.~\ref{fig:density_comparison} shows the result of adding only velocity kick heating to the previous calculation (dashed lines). Velocity kicks reduce density profiles relative to CDM for $r/R_{\mathrm{vir}}\lesssim 0.3$; this effect is stronger for lower-mass halos and at later times. For a $10^{10}~M_{\mathrm{\odot}}$ NFW halo with a median concentration at $z=0$, $v_{\mathrm{k}}/v_{\mathrm{e}}$ grows as $r/R_{\mathrm{vir}}$ increases and exceeds unity at $r/R_{\mathrm{vir}}\approx 0.2$; at higher redshifts, this scale moves to larger radii. This explains why the characteristic radius at which velocity kicks begin to significantly suppress DDM density profiles is $r/R_{\mathrm{vir}}\approx 0.2$ and is consistent with our prediction for the redshift evolution of $\rho_{\mathrm{DDM}}/\rho_{\mathrm{CDM}}$, which is most visible in the left panel of Fig.~\ref{fig:density_comparison}. Although $v_{\mathrm{k}}/v_{\mathrm{e}}$ continues to increase at even larger radii, $\sigma_{v}$ decreases in halos' outer regions, yielding a higher fraction of retained (and preferentially cold) particles. This competing effect explains why $\rho_{\mathrm{DDM}}/\rho_{\mathrm{CDM}}$ approaches unity for $r/R_{\mathrm{vir}}\gtrsim 0.3$. The resulting outer DDM density profiles can even exceed CDM as mass is pushed outward, as shown in the left panel of Fig.~\ref{fig:density_comparison}.

Finally, we add mass loss--induced DDM heating to the previous calculation (solid lines). For the $10^9~M_{\mathrm{\odot}}$ halo, this effect substantially lowers $\rho_{\mathrm{DDM}}/\rho_{\mathrm{CDM}}$ at all radii, and particularly in the inner regions where the $r^{-1}$ term in Eq.~(\ref{eq:delta_E_U_DDM}) dominates. Meanwhile, the $10^{10}~M_{\mathrm{\odot}}$ halo is less strongly affected. We conclude that DDM halos with $V_{\mathrm{max}}\gtrsim v_{\mathrm{k}}$ are mainly affected by velocity kicks, while lower-mass halos with $V_{\mathrm{max}}\lesssim v_{\mathrm{k}}$ are mainly impacted by mass loss--induced heating.

\section{Subhalo Populations}
\label{sec:subhalos}

The DDM effects discussed above also apply to subhalos, both before and after they fall into a larger host. After accretion, subhalos are additionally subject to tidal forces, which reduce the abundance of surviving DDM subhalos relative to CDM in a radially dependent fashion because cored DDM subhalos are stripped and disrupted more efficiently than cuspy CDM subhalos. Here, we study the impact of DDM on both subhalo mass functions and radial distributions.

\subsection{Setup}

We compare our subhalo population predictions to the cosmological N-body simulation results from Ref.~\cite{DES:2022doi} to validate our model, leaving a more detailed exploration of DDM parameter space to a future study. In particular, Ref.~\cite{DES:2022doi} presented cosmological DDM zoom-in simulations of a $\approx 10^{12}~M_{\mathrm{\odot}}$ host for $v_{\mathrm{k}}=20~\mathrm{km\ s}^{-1}$ (with $\tau = 10$, $20$, and $40~\mathrm{Gyr}$) and $v_{\mathrm{k}}=40~\mathrm{km\ s}^{-1}$ (with $\tau = 20$, $40$, and $80~\mathrm{Gyr}$); a subset of these simulations were presented in Ref.~\cite{Wang:2014ina}. The particle mass resolution is $\approx 2\times 10^5~M_{\mathrm{\odot}}$, and the Plummer-equivalent force softening length is $143~\mathrm{pc}$.

To model subhalo populations, we build merger trees using the same algorithm and halo concentration models described in Sec.~\ref{sec:isolated}. For CDM and each DDM model from Ref.~\cite{DES:2022doi}, we generate $100$ merger trees with a $z=0$ host mass of $10^{12}~M_{\mathrm{\odot}}$ and a merger tree mass resolution of $2\times 10^7~M_{\mathrm{\odot}}$, which is chosen to emulate the minimum number of $\approx 100$ particles needed to identify halos in the cosmological simulations. Building these merger trees determines subhalo properties at infall.

We then evolve subhalos in the time-evolving host potential using the orbital model from Ref.~\cite{Pullen:2014gna}. For CDM, we implement the best-fit tidal heating model from Ref.~\cite{Du:2024sbt}. For DDM, we set the second-order energy perturbation coefficient from Ref.~\cite{Du:2024sbt} to zero since this term is most relevant for cuspy halos. We also include the finite-resolution density profile model from Ref.~\cite{Benson:2022tzm}, which introduces small artificial cores in subhalos' centers, with a core size determined by the mass resolution and force softening of the cosmological simulation against which we compare. This model reproduces artificial disruption in cosmological CDM simulations. Including this model only appreciably affects our CDM predictions, since DDM core sizes are generally larger than those introduced by the finite-resolution effect.

We track stripped subhalos down to a bound mass threshold of $2\times 10^5~M_{\mathrm{\odot}}$, corresponding to the N-body simulation particle mass. Following the convergence thresholds used in Ref.~\cite{DES:2022doi}, we respectively apply cuts on present-day and peak maximum circular velocity of $V_{\mathrm{max}}>9~\mathrm{km\ s}^{-1}$ and $V_{\mathrm{peak}}>10~\mathrm{km\ s}^{-1}$, which effectively limit our analyses to subhalos with bound masses above $\approx 2\times 10^7~M_{\mathrm{\odot}}$ at $z=0$.

\begin{figure*}[t!]
 \centering
 \includegraphics[width=\textwidth]{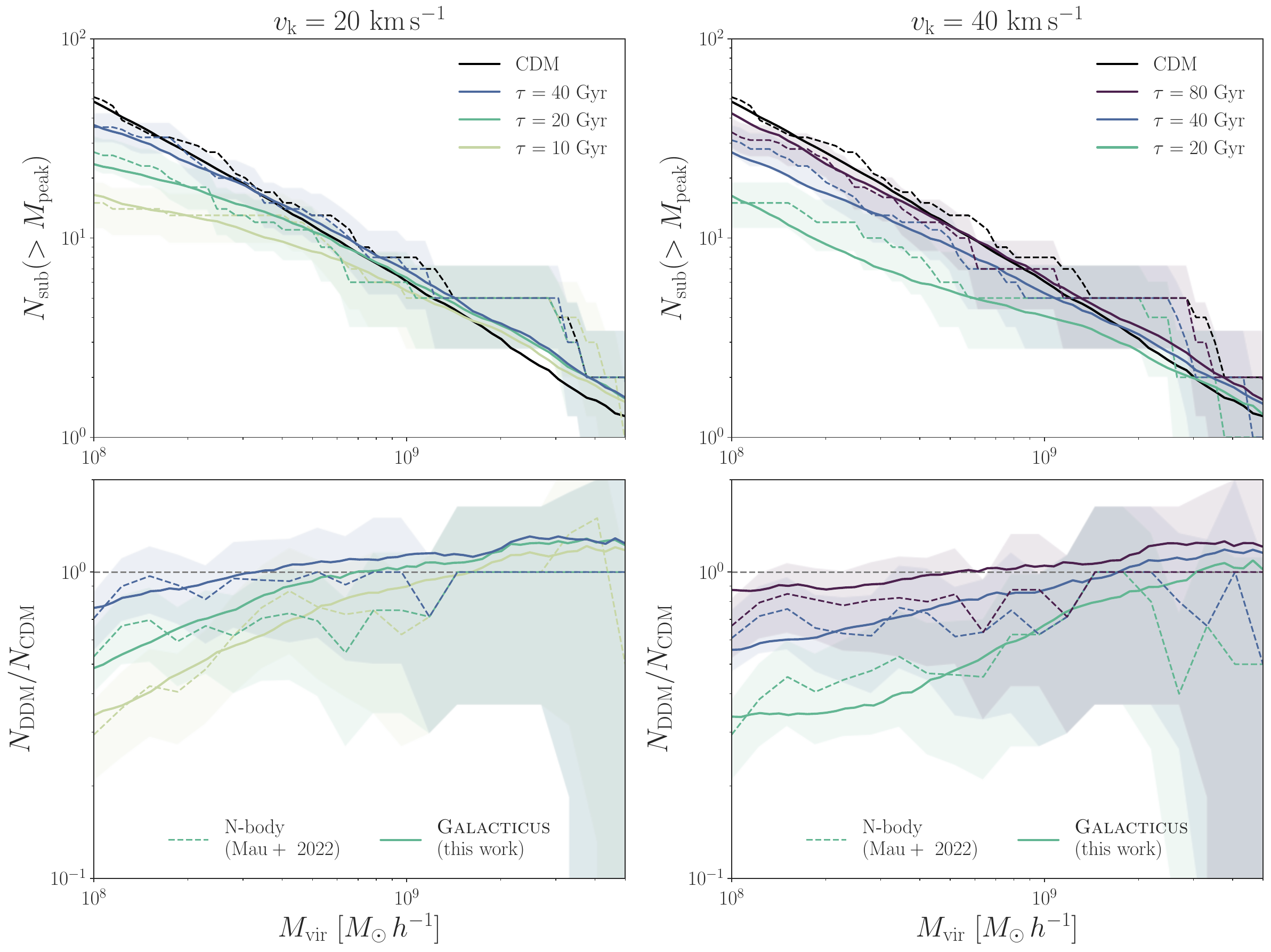}
 \caption{Cumulative subhalo mass functions (top) and DDM-to-CDM subhalo mass function ratios (bottom) for CDM (black) and DDM models with $v_{\mathrm{k}}=20$ (left) and $40~\mathrm{km\ s}^{-1}$ (right). Mean DDM results are shown for $\tau = 10$ (gold), $20$ (green), $40$ (blue), and $80~\mathrm{Gyr}$ (purple). Solid lines show our semianalytic \textsc{Galacticus} results, and dashed lines show the mean cosmological N-body DDM simulation results from Ref.~\cite{DES:2022doi}. Shaded bands show $68\%$ confidence interval Poisson uncertainties on the simulation measurements. In both cases, we only include subhalos with present-day maximum circular velocity $V_{\mathrm{max}}>9~\mathrm{km\ s}^{-1}$ and peak maximum circular velocity $V_{\mathrm{peak}}>10~\mathrm{km\ s}^{-1}$.} \label{fig:shmf}
\end{figure*}

\subsection{Subhalo mass functions}

Figure~\ref{fig:shmf} shows the resulting mean cumulative subhalo mass functions (SHMFs; top) and DDM-to-CDM SHMF ratios (bottom) from our analysis (solid) and the cosmological N-body simulations (dashed) measured using present-day virial mass.\footnote{Reference~\cite{DES:2022doi} analyzes SHMFs using peak (rather than present-day) subhalo mass. We measure SHMFs using present-day mass because \textsc{Galacticus} CDM predictions match recent zoom-in simulation results in this regard~\cite{Yang:2020aqk,Nadler:2022dvo}, while peak SHMFs are more sensitive to resolution cuts and halo finding algorithms~\cite{Mansfield:2023prs}.} We compare the mean results because only one N-body host is available. Our DDM predictions match the N-body simulations within the $68\%$ confidence interval Poisson uncertainties, shown by the shaded bands, across the entire range of well-resolved subhalo masses. The SHMF depends on both the density profiles of our subhalos at infall and their subsequent tidal evolution; the agreement between our predictions and the N-body simulations thus lends confidence to both our DDM and tidal evolution models.

The severity of DDM SHMF suppression is determined by both $v_{\mathrm{k}}$ and $\tau$. For $v_{\mathrm{k}}=20~\mathrm{km\ s}^{-1}$, we predict that $34\pm 12\%$, $49\pm 14\%$, and $76\pm 22\%$ of the total CDM subhalo population with $M_{\mathrm{vir}}>10^8~M_{\mathrm{\odot}}~h^{-1}$ survives for $\tau=10$, $20$, and $40~\mathrm{Gyr}$, respectively, where uncertainties represent the standard deviation across our $100$ merger trees. For $v_{\mathrm{k}}=40~\mathrm{km\ s}^{-1}$, these ratios become $34\pm 11\%$, $56\pm 21\%$, and $87\pm 31\%$ for $\tau=20$, $40$, and $80~\mathrm{Gyr}$, respectively. Similar to Ref.~\cite{DES:2022doi}, the total SHMF suppression we predict is consistent with being determined by $v_{\mathrm{k}}/\tau$; for example, SHMF suppression is similar for our $\{v_{\mathrm{k}}=20~\mathrm{km\ s}^{-1},\tau=10~\mathrm{Gyr}\}$ and $\{v_{\mathrm{k}}=40~\mathrm{km\ s}^{-1},\tau=20~\mathrm{Gyr}\}$ models. It will be interesting to study the origin of this scaling, since DDM effects on halo structure are nonlinear for sufficiently large $v_{\mathrm{k}}$ according to our results in Sec.~\ref{sec:isolated}.

For both $v_{\mathrm{k}}=20$ and $40~\mathrm{km\ s}^{-1}$, we predict a slight (but systematic) SHMF \emph{enhancement} relative to CDM for $M_{\mathrm{vir}}\gtrsim 2\times 10^9~M_{\mathrm{\odot}}~h^{-1}$. This may be due to less effective tidal stripping in cored DDM host halos. Note that our calculation does not capture ``core stalling,'' i.e., less efficient dynamical friction acting on massive, cored subhalos~\cite{Read:2006fq,Banik:2021jtr}, because our tidal evolution model relies on the Chandrasekhar dynamical friction formula~\cite{Chandrasekhar1943}. Directly comparing the tidal evolution of massive subhalos to the N-body results, which do not show a similar enhancement at the high-mass end of the SHMF, is thus a promising area for future work.

\begin{figure*}[t!]
 \centering
 \includegraphics[width=\textwidth]{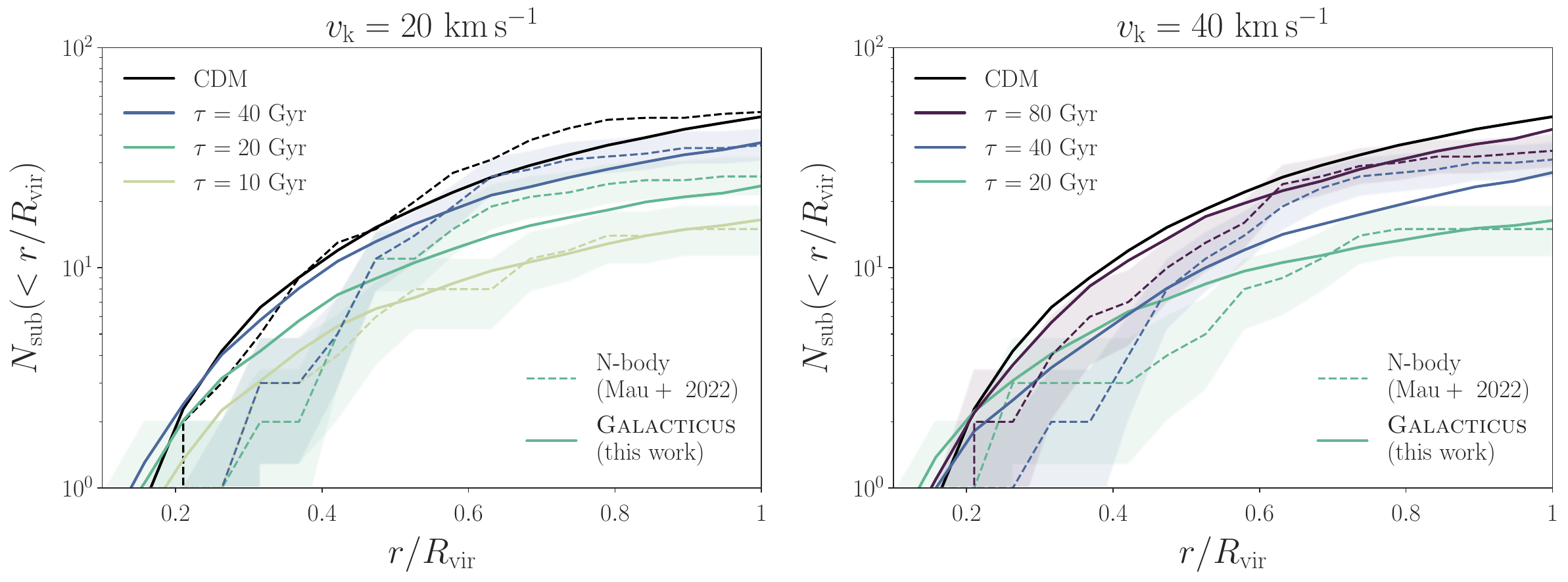}
 \caption{Same as Fig.~\ref{fig:shmf}, but showing cumulative subhalo radial distributions as a function of distance from the host center, in units of the host halo virial radius, for subhalos with $M_{\mathrm{vir}}>10^8~M_{\mathrm{\odot}}$.} \label{fig:shrf}
\end{figure*}

\subsection{Subhalo radial distribution}

Figure~\ref{fig:shrf} shows cumulative subhalo radial distributions (SHRFs) subject to the same $V_{\mathrm{max}}$ and $V_{\mathrm{peak}}$ cuts as above; we additionally apply a $M_{\mathrm{vir}}>10^8~M_{\mathrm{\odot}}~h^{-1}$ cut for direct comparison with the mass range shown in Fig.~\ref{fig:shmf}. We measure SHRFs in units of subhalo distance from the host center, normalized by the host's virial radius $R_{\mathrm{vir}}$, following Ref.~\cite{DES:2022doi}.

We find that \textsc{Galacticus} and N-body SHRFs are consistent at large radii ($r/R_{\mathrm{vir}}\gtrsim 0.5$) but that \textsc{Galacticus} predicts systematically higher subhalo abundances at smaller radii in both CDM and DDM. The magnitude of this discrepancy is consistent across all DDM models and is similar to previous comparisons between \textsc{Galacticus} and cosmological zoom-in simulations in CDM~\cite{Nadler:2022dvo}. We note that this SHRF discrepancy is mass dependent, and we refer the reader to Ref.~\cite{Benson:2022tzm} for further comparisons between \textsc{Galacticus} predictions and N-body subhalo populations.

We predict that the DDM-to-CDM subhalo mass function suppression is roughly independent of distance from the host center, consistent with the N-body simulation results from Ref.~\cite{DES:2022doi}. Interestingly, we find that $N_{\mathrm{DDM}}/N_{\mathrm{CDM}}$ slightly increases in the innermost regions ($r/R_{\mathrm{vir}}\lesssim 0.3$). One potential explanation is that DDM impacts low-mass subhalos sufficiently strongly such that they disrupt even at large distances from the host center, counteracting the fact that low-mass subhalos are less radially concentrated than high-mass subhalos overall. These effects warrant detailed follow-up studies.

\section{Summary and Discussion}
\label{sec:discussion}

We have presented a semianalytic model for DDM halos that is integrated into the \textsc{Galacticus} framework and publicly available at \url{http://github.com/galacticusorg/galacticus}. The model captures the effects of DDM mass loss and velocity kicks on halo profiles, with heating effects incorporated into a generalized adiabatic heating framework. This allows us to efficiently and accurately predict halo and subhalo density profile evolution as a function of the decay lifetime and kick velocity that parametrize the two-body DDM model.

Our main findings are as follows:

\begin{itemize}
    \item DDM halos' density and circular velocity profiles are suppressed relative to CDM. This suppression grows over time and is more severe for halos with $V_{\mathrm{max}}\lesssim v_{\mathrm{k}}$ (Fig.~\ref{fig:density_velocity}).
    \item The $R_{\mathrm{max}}$--$V_{\mathrm{max}}$ relation for DDM halos with $V_{\mathrm{max}}\lesssim v_{\mathrm{k}}$ bends away from CDM, toward smaller $V_{\mathrm{max}}$ and larger $R_{\mathrm{max}}$ (Fig.~\ref{fig:rmax_vmax}).
    \item The density profiles of DDM halos with $V_{\mathrm{max}}\gtrsim v_{\mathrm{k}}$ are mainly impacted by velocity kicks, while halos with $V_{\mathrm{max}}\lesssim v_{\mathrm{k}}$ are mainly impacted by mass loss--induced heating (Fig.~\ref{fig:density_comparison}).
    \item DDM subhalo mass functions are suppressed relative to CDM, particularly at the low-mass end ($M_{\mathrm{vir}}\lesssim 10^9~M_{\mathrm{\odot}}$), in agreement with previous cosmological DDM zoom-in simulations (Fig.~\ref{fig:shmf}).
    \item DDM does not significantly change the shape of the subhalo radial distribution relative to CDM, except in the innermost regions ($r/R_{\mathrm{vir}}\lesssim 0.3$; Fig.~\ref{fig:shrf}).
\end{itemize}

Our model assumes that the DDM heating energy is a fixed multiple of $GM/r$ within the shell-crossing radius. While we have shown that this assumption yields accurate DDM predictions for both isolated halos and subhalo populations relative to N-body simulations, it also introduces a sharp feature in our halo profile predictions at the (time-dependent) shell-crossing radius. This feature starts at $r/R_{\mathrm{vir}}\approx 0.2$, where the DDM kick velocity $v_{\mathrm{k}}$ becomes comparable to the escape velocity $v_{\mathrm{e}}$, and moves outward over time. This feature is not visible in DDM halo profiles from N-body simulations (e.g., from Ref.~\cite{Peter:2010jy}), which may be due to anisotropic accretion in a cosmological environment. It will be important to study this effect in follow-up work.

We have also assumed that tidal and DDM heating are independent. This assumption could be tested, and (if necessary) a better model constructed, using high-resolution simulations of DDM subhalos (e.g., following the approach of Ref.~\cite{Du:2024sbt}). Such tests would help clarify our finding that SHMF suppression is consistent with being determined by $v_{\mathrm{k}}/\tau$, which may only hold over a limited range of DDM parameter space.

\textsc{Galacticus} has recently been used to model high-redshift galaxy populations~\cite{Nadler:2022kmy,Driskell241011680}, dwarf satellite galaxies~\cite{Weerasooriya,Ahvazi}, stellar stream impacts~\cite{Menker:2024koc}, strong lensing substructure~\cite{Keeley:2024brx}, and subhalo evolution~\cite{Du:2024sbt}. Our DDM model can easily be tested in any of these settings. It is particularly compelling to integrate our model into the MW satellite forward-modeling framework used in Ref.~\cite{DES:2022doi}, since it allows constraints to be evaluated at kick velocity values other than $v_{\mathrm{k}}=20$ and $40~\mathrm{km\ s}^{-1}$. Our work thus enables a comprehensive scan over DDM parameter space, which we look forward to pursuing.


\section*{Acknowledgements}

We are very grateful to Juan Quiroz for contributing to the initial stages of this project through the Carnegie Astrophysics Summer Student Internship Program. We thank the anonymous referee for a constructive report, Xiaolong Du for comments on the manuscript, George Fuller for helpful discussions, and Sidney Mau for comments on the manuscript and providing the DDM simulation data from Ref.~\cite{DES:2022doi}. Computing resources used in this work were made available by a generous grant from the Ahmanson Foundation.

\section*{Data Availability}

The data that support the findings of this article are openly available~\cite{galacticus_commit_88bf30e}.

\bibliography{references}


\appendix

\begin{figure*}[ht!]
\centering
\includegraphics[width=0.49\textwidth]{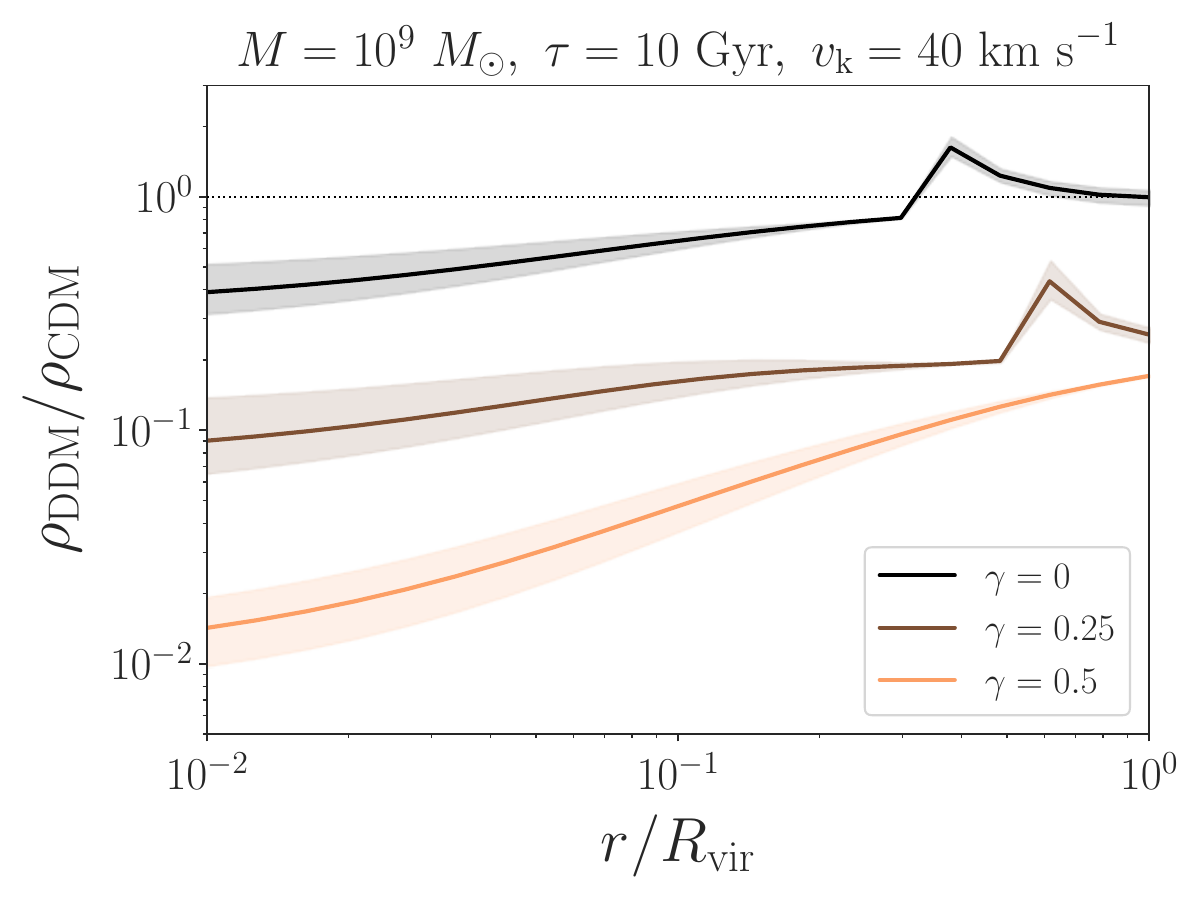}
\includegraphics[width=0.49\textwidth]{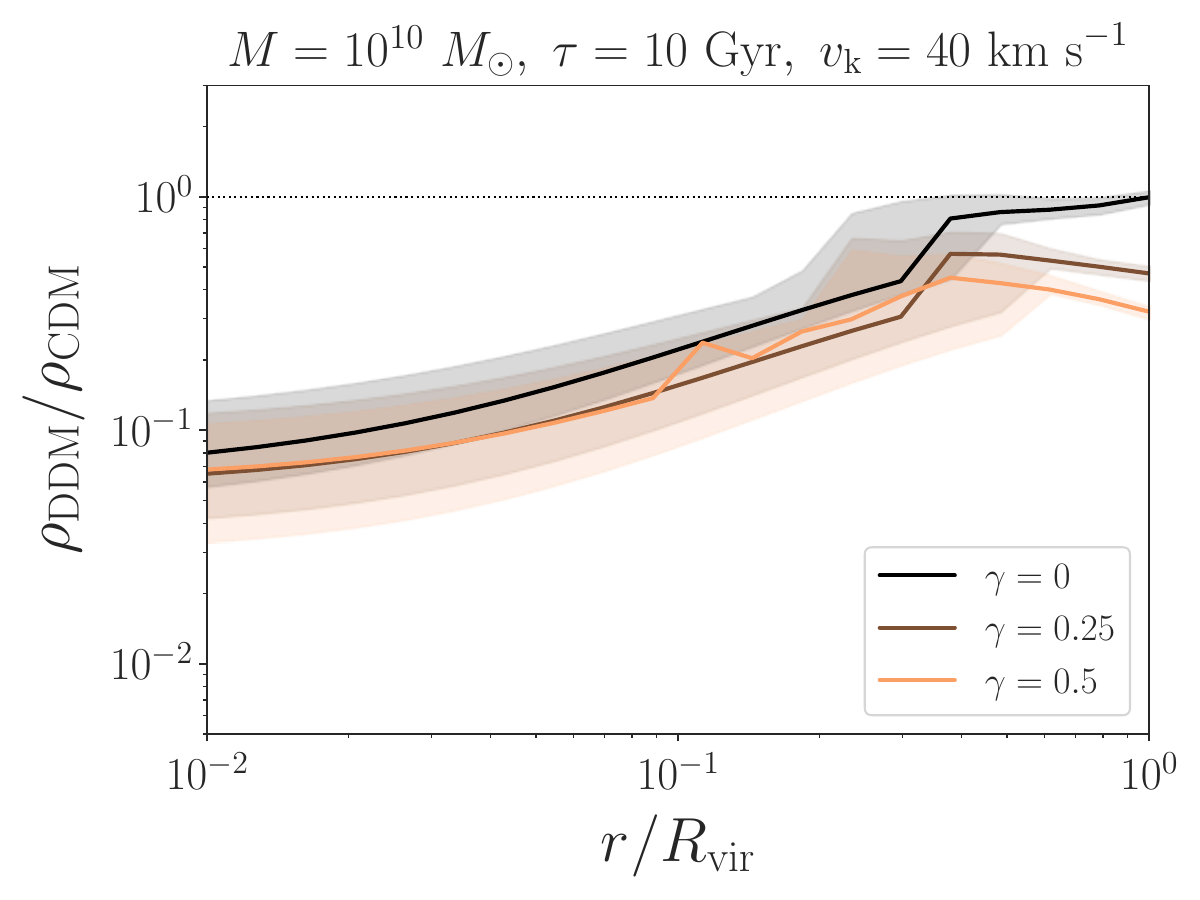}
\vspace{-2mm}
    \caption{Same as Fig.~\ref{fig:density_comparison}, but comparing DDM density profile ratios for $\gamma=0$, $0.25$, and $0.5$, from darkest to lightest.}
    \label{fig:density_gamma_comparison}
\end{figure*}

\section{Varying Mass Loss Heating Efficiency}
\label{sec:gamma}

Our fiducial results assume a mass loss efficiency parameter of $\gamma=0.5$. In Fig.~\ref{fig:density_comparison}, we compared this choice to an extreme scenario without mass loss heating. In this Appendix, we explore the effects of varying $\gamma$ in more detail, study its degeneracy with DDM parameters, and comment on how it can be determined in future work.

Fig.~\ref{fig:density_gamma_comparison} repeats our calculation from Fig.~\ref{fig:density_comparison} including an additional model with $\gamma=0.25$. In the $10^9~M_{\mathrm{\odot}}$ halo, the suppression of the inner density profile relative to CDM increases smoothly as $\gamma$ grows. Meanwhile, the shell-crossing feature in the outer density profile moves toward large radii, and moves past the virial radius for $\gamma=0.5$. The effect of varying $\gamma$ is more subtle for the $10^{10}~M_{\mathrm{\odot}}$ halo, although there are still noticeable differences in the outer density profile as a function of  in this case. We do not show results for $\gamma>0.5$ because mass shells can become unbound in this regime; a more sophisticated shell-crossing model will be needed to capture such models accurately.

Thus, our numerical parameter $\gamma$ and the physical DDM parameters $\tau$ and $V_{\mathrm{kick}}$ have partially degenerate effects on the density profiles of low-mass halos. This degeneracy can in principle be broken by subhalo population statistics. For example, our predictions using $\gamma=0.5$ agree with SHMFs from DDM N-body simulations according to Fig.~\ref{fig:shmf}. While calibrating $\gamma$ using the N-body simulation data is beyond the scope of this work and would require a more sophisticated treatment of both simulation and modeling uncertainties, this result already indicates that our fiducial model with $\gamma=0.5$ is reasonable.

\end{document}